\documentclass[12pt]{article}
\usepackage{cite}
\begin{document}

\author{I.M. Sokolov$^{1}$, A. Blumen$^{1}$ and J. Klafter$^{2}$ \\
%EndAName
$^{1}$Herman-Herder-Str.3, D-79104 Freiburg im Breisgau, Germany\\
$^{2}$School of Chemistry, Tel Aviv University, 69978 Tel Aviv, Israel}
\title{Dynamics of Annealed Systems under External Fields: CTRW and the Fractional
Fokker-Planck Equations. }
\date{\today}
\maketitle

\begin{abstract}
We consider the linear response of a system modelled by continuous-time
random walks (CTRW) to an external field pulse of rectangular shape. We
calculate the corresponding response function explicitely and show that it
exhibits aging, i.e. that it is not translationally invariant in the
time-domain. This result differs from that of systems which behave according
to fractional Fokker-Planck equations.

\end{abstract}

The response dynamics of several physical systems to external fields is
well-described by the continuous-time random walk (CTRW) model introduced by
Montroll and Weiss \cite{MoWei}. This model was extremely successful in the
explanation of dispersive transport in amorphous semiconductors \cite{ScheM}%
, see \cite{HausKehr,BouGeor} for reviews. Recently it has been shown that
in time-independent fields CTRW dynamics is sometimes rendered well through
fractional calculus relations such as fractional Fokker-Planck equations
(FFPE) \cite{MeKla}. Now FFPEs have considerable mathematical advantages
when compared to discrete stochastic schemes such as CTRW, since they allow
to incorporate readily different initial and boundary conditions. In
time-independent fields the FFPE expressions follow from those of CTRWs
using the Kramers-Moyal expansion \cite{BMK}. Here we show that care is to
be taken when treating \textit{time-dependent} external fields. We
demonstrate that for these the linear response obtained in CTRW and in FFPE
schemes may differ considerably; we display this here by taking exemplarily
an external field which is switched on and off, i.e. which is of rectangular
shape in time. In this case the CTRW-process displays aging, i.e. the
response depends explicitly on the delay between the time of measurement and
the time at which the system was prepared in a given state; hence the
response reveals an essentially non-equilibrium process. On the other hand,
FFPE follows the usual linear response behavior of systems close to
equilibrium \cite{Sokolov1} and is expected to be applicable in the cases of
polymers \cite{Schiessel} and rough interfaces \cite{Liu}. The differences
between CTRW and FFPE turn out to be very pronounced in the time just after
switching off the external field.

As is usual in CTRW we envisage an ensemble of noninteracting particles
which may be influenced by external fields (say, the particles are charged).
The particles follow then CTRWs, i.e. sequences of jumps. The time intervals 
$t_{i}$ between the jumps are uncorrelated. Of interest are waiting times
which follow power-law distributions, i.e. 
\begin{equation}
\psi (t)=\gamma /(1+t/\tau _{0})^{1+\gamma },\qquad \mbox{with }0<\gamma <1.
\label{wtd1}
\end{equation}
The physical motivation for such $\psi (t)$-forms may be rationalized using
random traps whose energy distribution is exponential \cite{BKZ}. In what
follows we put $\tau _{0}=1$ and work in dimensionless time units.

A basic quantity in the CTRW formalism is $\chi _{n}(t)$, the probability to
make exactly $n$ steps up to time $t$. In the standard, decoupled CTRW
picture (in which the spatial transition probabilities between the lattice
sites are independent of the waiting-times) the probability distribution $P(%
\mathbf{r},t)$ of finding a particle at $\mathbf{r}$ at time $t$ given that
is started at $\mathbf{0}$ at time 0 obeys \cite{Blumofen} 
\begin{equation}
P(\mathbf{r},t)=\sum_{i=0}^{\infty }P_{n}(\mathbf{r})\chi _{n}(t),
\label{Green}
\end{equation}
where $P_{n}(\mathbf{r})$ is the probability to reach $\mathbf{r}$ from $%
\mathbf{0}$ in $n$ steps. We note that Eq.(\ref{Green}) is the starting
point for the derivation of FFPE in Ref.\cite{BMK}.

Using Eq.(\ref{Green}) several important relations follow: Thus the mean
particle's displacement equals $\overline{\mathbf{X}(t)}=\sum_{n=0}^{\infty }%
\overline{\mathbf{X}_{n}}\chi _{n}(t)$, where $\overline{\mathbf{X}_{n}}$ is
the mean displacement after $n$ steps. In a weak homogeneous and
time-independent external field $\mathbf{E}$, $\overline{\mathbf{X}_{n}}$ is
proportional to $\mathbf{E}$ and to $n$. In particular, if only jumps
between nearest neighbor (nn) sites are allowed, the mean displacement per
step is 
\begin{equation}
\mathbf{x}=\sum_{i}\mathbf{r}_{i}\frac{\left( \mathbf{r}_{i}\cdot \mathbf{E}%
\right) }{k_{B}T}=\mathbf{\hat{\mu}E,}  \label{Displ}
\end{equation}
where the sum runs over all nn vectors, $k_{B}$ is the Boltzmann constant, $%
T $ is the temperature and $\mathbf{\hat{\mu}}$ is the mobility tensor. In
highly symmetrical lattices $\mathbf{x\parallel E}$ for any field direction
and Eq.(\ref{Displ}) reduces to a scalar relation, $\mathbf{x}=\hat{\mu}%
\mathbf{E}$. Furthermore, since $\overline{\mathbf{X}_{n}}=n\mathbf{x=}n\hat{%
\mu}\mathbf{E}$: 
\begin{equation}
\overline{X(t)}=\sum_{i=0}^{\infty }xn\chi _{n}(t)=xN(t)=\hat{\mu}EN(t),
\label{Xmean}
\end{equation}
where $N(t)=\sum_{i=0}^{\infty }n\chi _{n}(t)$ is the mean number of steps
performed up to time $t$.

In what follows we restrict ourselves to isotropic lattices and use a scalar
notation. Typically, the evolution of $N(t)$ depends now on the initial
conditions \cite{HausKehr}. Thus, if we are interested in the number of
steps performed from the moment $t=0$ in which the system was initially
prepared (this is the usual situation in photoconductivity), all the
intervals between the steps are governed by the same $\psi (t)$. Then $%
N(\lambda )$, the Laplace-transform of $N(t)$, reads 
\begin{equation}
\tilde{N}(\lambda )=\frac{\tilde{\psi}(\lambda )}{\lambda \left[ 1-\tilde{%
\psi}(\lambda )\right] }  \label{Nmean}
\end{equation}
as is immediately evident from the expression for $\chi _{n}(\lambda )=\psi
(\lambda )^{n}\left[ 1-\psi (\lambda )\right] /\lambda $, Ref.\cite{BKZ}.
The Laplace transform of $\psi (t)$, Eq.(\ref{wtd1}), for small $\lambda $
is known to be $\tilde{\psi}(\lambda )=1-\lambda ^{\gamma }\Gamma (1-\gamma
) $ \cite{Shlesinger}. From Eq.(\ref{Nmean}) we now have 
\begin{equation}
\tilde{N}(\lambda )\simeq \frac{1}{\Gamma (1-\gamma )}\lambda ^{-(\gamma +1)}
\end{equation}
The inverse Laplace-transform of this expression reads: 
\begin{equation}
N(t)=\frac{1}{\Gamma (1+\gamma )\Gamma (1-\gamma )}t^{\gamma }=\frac{\sin
\pi \gamma }{\pi \gamma }t^{\gamma }  \label{N1}
\end{equation}

Consider now the case that $E$ is switched on at the time $t_{w}$ after the
initial preparation of the system. In this case one has $x=0$ for $t<t_{w}$,
so that only the number of steps performed after $t_{w}$ matters. This
situation was discussed in Ref.\cite{Tunaley} (see \cite{HausKehr} for a
review). Now, the mean number of steps performed during the observation time 
$\tau =t-t_{w}$, i.e. between $t_{w}$ and $t$, is 
\begin{eqnarray}
N(\tau ,t_{w}) &=&N(t)-N(t_{w})=\frac{\sin \pi \gamma }{\pi \gamma }\left(
t^{\gamma }-t_{w}^{\gamma }\right) =  \nonumber \\
&=&\frac{\sin \pi \gamma }{\pi \gamma }\left[ (\tau +t_{w})^{\gamma
}-t_{w}^{\gamma }\right] .  \label{result}
\end{eqnarray}
Eq.(\ref{result}) fits very well the results of numerical simulations, see
Figs.1 and 2, where the simulated points are obtained from $10^{5}$
realizations of the random motion of walkers whose waiting-time
distributions follow Eq.(\ref{wtd1}). The results of the simulations are
marked by symbols, and the results of Eq.(\ref{result}) are shown as lines.
The parameters of Fig.1 are $\gamma =0.5$ and $t_{w}$ varies from 300 to
30000. In Fig. 2 we show the results for $\gamma =0.3,0.5$ and 0.75 as
functions of $\tau $ for a fixed value of $t_{w}=1000$. Note that no fit
parameters are used.

Let us now return to the motion in the field $E$. The particles'
displacements grows as $\hat{\mu}EN(\tau )$ and the related current is 
\begin{equation}
j=\hat{\mu}E\frac{d}{d\tau }N(\tau )=\frac{\sin \pi \gamma }{\pi }\hat{\mu}%
E(\tau +t_{w})^{\gamma -1}.  \label{resCurr}
\end{equation}
Note that the overall response function $\sigma (\tau ,t_{w})$ defined
through $j=\sigma (\tau ,t_{w})E$ is given by $\sigma (\tau ,t_{w})=\left(
\pi ^{-1}\sin \pi \gamma \right) \hat{\mu}t_{w}^{\gamma -1}(1+\tau
/t_{w})^{\gamma -1}$, and shows simple scaling with respect to its two time
variables: $\sigma (\tau ,t_{w})=t_{w}^{\gamma -1}F(t/t_{w})$. Thus in the
limit of short observation times, $\tau \ll t_{w}$, the response function $%
\sigma (\tau ,t_{w})=\left( \pi ^{-1}\sin \pi \gamma \right) \hat{\mu}%
t_{w}^{\gamma -1}$ is $\tau $-independent and hence describes an Ohmic
transport. However, the value of the conductivity decays with the delay
time. Systems in which the response to an external agent depends explicitly
on the delay between preparation time and acting (measurement) time are
referred to as aging systems. This kind of behavior was found to be very
pronounced in CTRWs with $0<\gamma <1$ \cite{Feigelman,Bouchaud,Montus,Maas}%
. For $\tau \gg t_{w}$, on the other hand, $j=\left( \pi ^{-1}\sin \pi
\gamma \right) \hat{\mu}E\tau ^{\gamma -1}$; the current is hence dispersive
and independent of $t_{w}$.

Let us now consider the response of the system to a field switched on at
time $t_{w}$ and switched off at time $t_{z}$. Note that when the field is
switched off, the directed component of motion ceases immediately: there is
no afteraction. Hence the current is 
\begin{equation}
j(t)=\left\{ 
\begin{array}{ll}
0 & t<t_{w} \\ 
\left( \pi ^{-1}\sin \pi \gamma \right) \hat{\mu}Et^{\gamma -1} & t_{w}\leq
t\leq t_{z} \\ 
0 & t>t_{z}
\end{array}
\right.  \label{agedresp}
\end{equation}
which is a causal response concentrated on the time interval in which the
field acts.

We now continue by discussing our findings in relation to a recently
introduced approach to slow relaxation in time-independent external fields,
based on fractional Fokker-Planck equations (FFPEs) \cite{MeKla,Metz1} or on
fractional Master equations \cite{AH}. The FFPE describing subdiffusive
behavior in an external field reads:

\begin{equation}
\frac{\partial }{\partial t}P(x,t)=\,_{0}D_{t}^{1-\gamma }\mathcal{L}%
_{FP}P(x,t),  \label{subdiff}
\end{equation}
where $P(x,t)$ is the pdf to find a particle (walker) at point $x$ at time $%
t $. In Eq.(\ref{subdiff}) $_{0}D_{t}^{1-\gamma }$ is the fractional
Riemann-Liouville operator ($0<\gamma <1$) defined through \cite{MeKla} 
\begin{equation}
_{0}D_{t}^{1-\gamma }Z(t)=\frac{1}{\Gamma (\gamma )}\frac{\partial }{%
\partial t}\int_{0}^{t}dt^{\prime }\frac{Z(t^{\prime })}{(t-t^{\prime
})^{1-\gamma }}.  \label{RiLi}
\end{equation}
(in which $t=0$ can be associated with the time at which the system was
prepared). Furthermore, in Eq.(\ref{subdiff}) $\mathcal{L}_{FP}$ is the
Fokker-Planck operator 
\begin{equation}
\mathcal{L}_{FP}=K\frac{{\partial }^{2}}{{\partial }x^{2}}-\mu \frac{%
\partial }{\partial x}f(x)=K\frac{{\partial }^{2}}{{\partial }x^{2}}-\mu E%
\frac{\partial }{\partial x}.  \label{FPOp}
\end{equation}
The r.h.s. of Eq.(\ref{FPOp}) holds, since in our case the acting force $%
f(x)=-\partial U/\partial x$ is homogeneous, $x$-independent and of
magnitude $E$. The fractional Fokker-Planck equation, Eq.(\ref{subdiff}),
has turned out to be useful in describing many phenomena connected with
anomalous diffusion or relaxation patterns \cite{MeKla}. We note that Eq.(%
\ref{subdiff}) can be derived from the decoupled scheme, Eq.(\ref{Green})
using the Kramers-Moyal expansion, Ref.\cite{BMK}, provided the field is
time-independent. The free relaxation properties of the CTRW system are
reproduced by Eq.(\ref{subdiff}) and can be expressed in terms of the
Mittag-Leffler functions \cite{MeKla,AH}.

We can now make connection to $\overline{X(t)}$ of Eq.(\ref{Xmean}) by
remarking that 
\begin{equation}
\overline{X(t)}=\int_{-\infty }^{\infty }xP(x,t)dx.
\end{equation}
Then multiplying Eq.(\ref{subdiff}) by $x$ and integrating it over the whole
axis we get (parallel to a simple diffusion equation): 
\begin{equation}
\frac{\partial }{\partial t}\int_{-\infty }^{\infty
}xP(x,t)dx=\,_{0}D_{t}^{1-\gamma }\left( K\int_{-\infty }^{\infty }x\frac{{%
\partial }^{2}}{{\partial }x^{2}}P(x,t)dx-\mu E(t)\int_{-\infty }^{\infty }x%
\frac{\partial }{\partial x}P(x,t)dx\right) .  \label{part}
\end{equation}
The left hand side of Eq.(\ref{part}) is nothing but $\frac{d}{dt}\overline{%
X(t)}$, whereas the right hand side can be simplified by integration by
parts. Using now that $P(x,t)$ and its derivatives with respect to $x$
vanish at infinity (as is reasonable for our models), the first integral
vanishes and the second one is unity. Hence 
\begin{equation}
\frac{d}{dt}\overline{X(t)}=\,_{0}D_{t}^{1-\gamma }\mu E(t).
\end{equation}
Concentrating on the temporal dependence, we get in the case that the
external field is switched on at $t=t_{w}$ and off at $t=t_{z}$, i.e. $%
E(t)=E\theta (t^{\prime }-t_{w})\theta (t_{z}-t^{\prime })$ that 
\begin{equation}
j(t)=\frac{d}{dt}\overline{X(t)}=\frac{\mu }{\Gamma (\gamma )}E\frac{d}{dt}%
\int_{0}^{t}dt^{\prime }\frac{\theta (t^{\prime }-t_{w})\theta
(t_{z}-t^{\prime })}{(t-t^{\prime })^{1-\gamma }}  \label{LiRe2}
\end{equation}
Evaluating the r.h.s. of Eq.(16) we get: 
\begin{equation}
j(t)=\left\{ 
\begin{array}{ll}
0 & t<t_{w} \\ 
\left[ \mu /\Gamma (\gamma )\right] E(t-t_{w})^{\gamma -1} & t_{w}\leq t\leq
t_{z} \\ 
\left[ \mu /\Gamma (\gamma )\right] E\left[ \left( t-t_{w}\right) ^{\gamma
-1}-\left( t-t_{z}\right) ^{\gamma -1}\right] & t>t_{z}
\end{array}
\right.  \label{Fracresp}
\end{equation}
which shows a functional dependence on the external field that strongly
differs from Eq.(\ref{agedresp}) as soon as $t>t_{w}$.

Equations (\ref{agedresp}) and (\ref{Fracresp}) coincide only if one
supposes $\mu /\Gamma (\gamma )=\hat{\mu}\pi ^{-1}\sin \pi \gamma $ and
takes $t_{w}=0$, $t_{z}\rightarrow \infty $; this limit parallels the
findings of Ref. \cite{BMK}. In general, however, the difference is large:
Thus, Eq.(\ref{agedresp}) describes a response concentrated on the
time-interval of the field-action, $t_{w}\leq t\leq t_{z}$: no afteraction
effects are seen. The current never changes sign and has finite jumps at $%
t=t_{w}$ and $t=t_{z}$. On the other hand, Eq.(\ref{Fracresp}) shows
considerable afteraction: the current does not vanish for $t>t_{z}$.
Moreover, the current diverges at $t=t_{w}$ and $t=t_{z}$ and changes its
sign from positive to negative at $t=t_{z}$. Moreover, the overall response
described by Eq.(\ref{Fracresp}) is invariant under time translation, i.e.
depends only on the differences $t-t_{w}$ and $t-t_{z}$, which is not the
case for CTRW, Eq.(\ref{agedresp}).

The structure of the response, Eq.(\ref{Fracresp}) derives from linear
response close to equilibrium \cite{Sokolov1,Schiessel,Metz1}, where in
general the response $I(t)$ and the field $E(t)$ are related through the
causal linear integral operator 
\begin{equation}
I(t)=\int_{t_{0}}^{t}\phi (t-t^{\prime })E(t^{\prime })dt^{\prime }
\label{LiRe}
\end{equation}
which does not account for aging effects. Memory effects in CTRW do not
arise from a memory kernel, as in FFPE, but from temporal \textit{%
subordination} (see Refs.\cite{BKZ,Sokolov}), which is an integral construct
different from the \textit{convolution} of Eq.(\ref{LiRe}). Note that
systems whose dynamics is governed by subdiffusive CTRW (i.e. for which $%
\gamma <1$) show aging; thus measuring the system's response to a pulsed
field can be instrumental in determining whether the system obeys CTRW, Eq.(%
\ref{agedresp}) or FFPE, Eq.(\ref{Fracresp}).

Let us summarize our findings. We have considered the linear response of
systems governed by CTRW and by FFPE dynamics to an external field switched
on at $t=t_{w}$. We showed that for CTRW the response is initially ohmic and
that it becomes dispersive after times comparable to the waiting time $t_{w}$%
. We also studied the CTRW response to a rectangular field pulse of finite
duration, switched on at $t=t_{w}$ and off at $t=t_{z}$. The form of this
response (retardation effects after switching the field on, but absence of
afteraction after switching it off) shows a behavior which differs
considerably from that displayed by systems obeying FFPE.

\begin{center}
{\Huge Acknowledgments}

\bigskip
\end{center}

The authors gratefully acknowledge financial support by the GIF, by the
Deutsche Forschungsgemeinschaft and by the Fonds der Chemischen
Industrie.

\newpage

\begin{center}
{\Huge Captions}
\end{center}

\textbf{Fig.1.} Mean number of steps $N(t)$ performed in time $t$, displayed
in double-logarithmic scales, for CTRW with a waiting-time distribution,
Eq.(1) with $\gamma =0.5$. Here the triangles, circles, diamonds, squares
and crosses indicate the $t_{w}$-values 300, 1000, 3000, 10000 and 30000,
respectively. The full lines reproduce Eq.(\ref{result}). The dotted line
has the slope 1, the dashed line has the slope 1/2.

\bigskip

\textbf{Fig.2.} Aging effects, to be seen from the dependence of $N$ on $%
t_{w}$ at $t=1000$. The CTRWs waiting time distributions obey Eq.(1): we
consider three different values of $\gamma $; namely $\gamma =0.75$
(circles), $\gamma =0.5$ (squares) and $\gamma =0.3$ (triangles). Note the
double-logarithmic scales. The full lines again reproduce Eq.(\ref{result}).

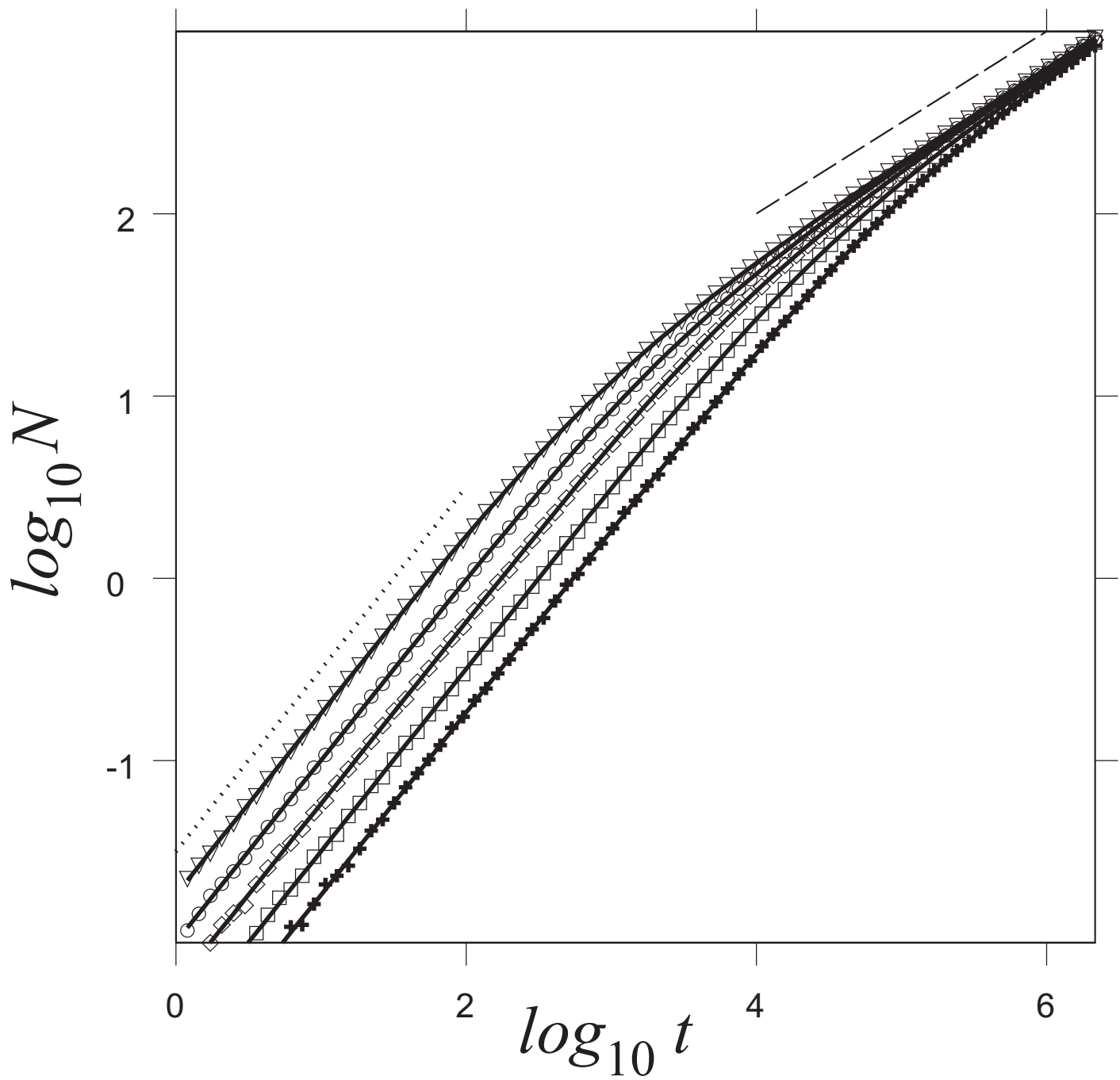
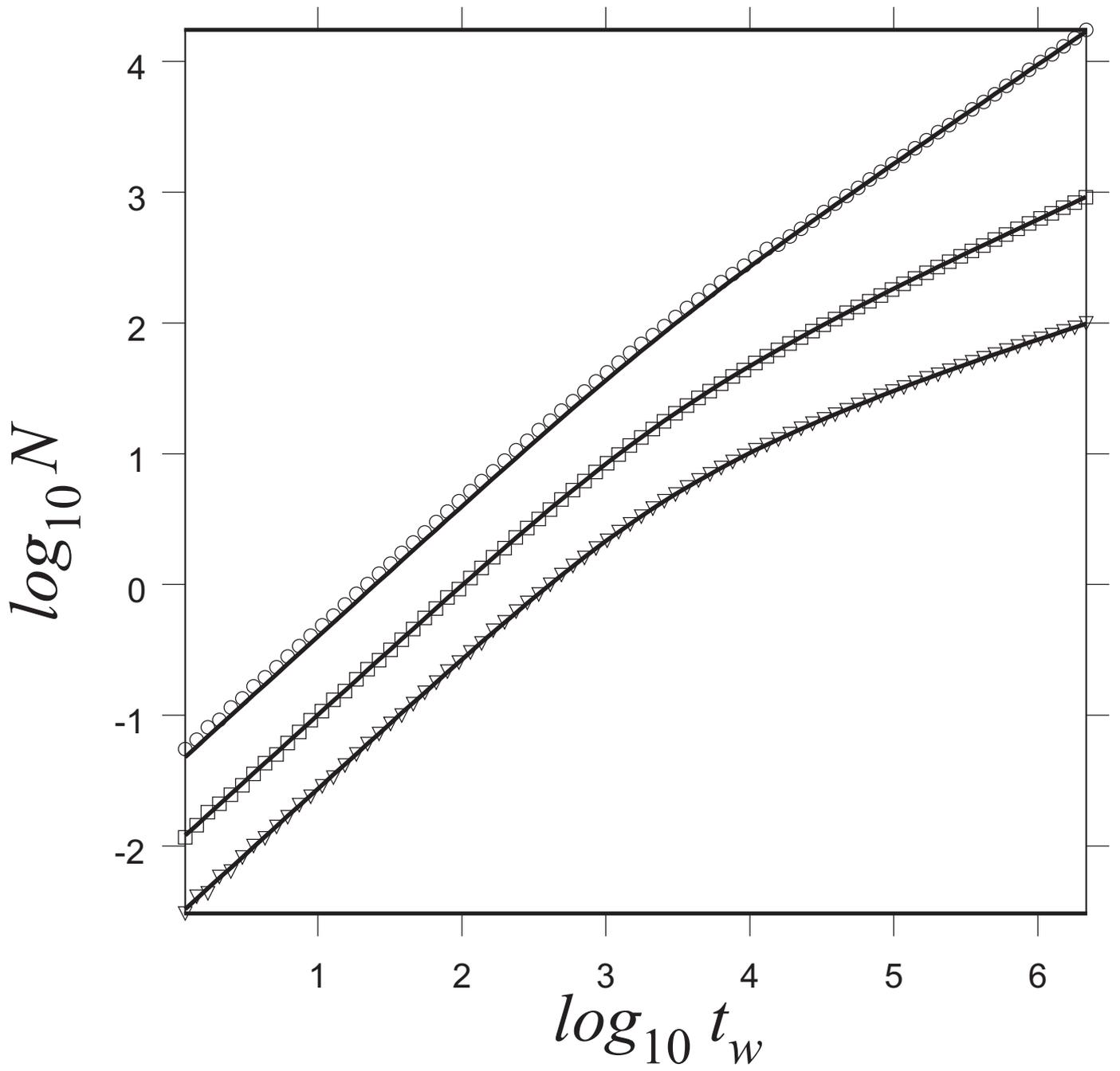
\end{document}